\documentclass[runningheads]{llncs}
\usepackage{amsmath}
\usepackage{verbatim}
\usepackage{ifthen}
\usepackage{url}
\usepackage{xspace}
\usepackage{subcaption}
\usepackage{calrsfs}
\usepackage{pifont}
\usepackage{amsfonts}
\usepackage{caption}
\usepackage{booktabs}
\usepackage{multirow}

\usepackage[hide]{cnote}
\usepackage[linesnumbered, ruled]{algorithm2e}

\usepackage{tikz}
\usetikzlibrary{shadows,shadings,shapes.symbols,positioning}

\tikzset{
diagonal fill/.style 2 args={fill=#2, path picture={
\fill[#1, sharp corners] (path picture bounding box.south west) -|
                         (path picture bounding box.north east) -- cycle;}},
reversed diagonal fill/.style 2 args={fill=#2, path picture={
\fill[#1, sharp corners] (path picture bounding box.north west) |- 
                         (path picture bounding box.south east) -- cycle;}}
}

\newcommand{\squishlist}{
 \begin{list}{$\bullet$}
  {  \setlength{\itemsep}{0pt}
     \setlength{\parsep}{3pt}
     \setlength{\topsep}{3pt}
     \setlength{\partopsep}{0pt}
     \setlength{\leftmargin}{2em}
     \setlength{\labelwidth}{1.5em}
     \setlength{\labelsep}{0.5em}
} }
\newcommand{\squishlisttight}{
 \begin{list}{$\bullet$}
  { \setlength{\itemsep}{0pt}
    \setlength{\parsep}{0pt}
    \setlength{\topsep}{0pt}
    \setlength{\partopsep}{0pt}
    \setlength{\leftmargin}{2em}
    \setlength{\labelwidth}{1.5em}
    \setlength{\labelsep}{0.5em}
} }

\newcommand{\squishdesc}{
 \begin{list}{}
  {  \setlength{\itemsep}{0pt}
     \setlength{\parsep}{3pt}
     \setlength{\topsep}{3pt}
     \setlength{\partopsep}{0pt}
     \setlength{\leftmargin}{1em}
     \setlength{\labelwidth}{1.5em}
     \setlength{\labelsep}{0.5em}
} }

\newcommand{\squishend}{
  \end{list}
}

\newcommand{\spara}[1]{\smallskip\noindent{\bf #1}}
\newcommand{\mpara}[1]{\medskip\noindent{\bf #1}}

\newcommand{\fpr}[1]{\mathopen{}\left(#1\right)}

\newcommand{\dispfunc}[2]{%
  \ensuremath{%
    \ifthenelse{\equal{\noexpand#2}{}}%
	     {#1}%
		      {{#1}\fpr{#2}}}}

\newcommand{\reals}{\ensuremath{\mathbb{R}}\xspace}
\newcommand{\realsnn}{\ensuremath{\reals_{+}}\xspace}

\DeclareMathAlphabet{\pazocal}{OMS}{zplm}{m}{n}

\newcommand{\bigO}{\ensuremath{\mathcal{O}}\xspace}
\newcommand{\np}{\ensuremath{\mathbf{NP}}\xspace}
\newcommand{\p}{\ensuremath{\mathbf{P}}\xspace}
\newcommand{\w}{\ensuremath{\mathbf{W}}\xspace}

\newcommand{\pc}[1]{{\pazocal{#1}}}
\newcommand{\mc}[1]{{\mathcal{#1}}}

\newcommand{\vertexcover}{\ensuremath{\text{\sc Vertex\-Cover}}\xspace}

\newcommand{\matmedian}{\ensuremath{\text{\sc Matroid\-Med\-ian}}\xspace}
\newcommand{\rbmedian}{\ensuremath{rb\text{\sc-Median}}\xspace}

\newcommand{\dominatingset}{\ensuremath{\text{\sc Dom\-Set}}\xspace}

\newcommand{\divkmedian}{\ensuremath{\text{\sc Div-}k\text{\sc-Median}}\xspace}
\newcommand{\divkmediannif}{\ensuremath{\text{\sc Div-}k\text{\sc-Median}_{\emptyset}}\xspace}

\newcommand{\lszero}{\ensuremath{\small\text{\sf LS-0}}\xspace}
\newcommand{\lsone}{\ensuremath{\small\text{\sf LS-1}}\xspace}
\newcommand{\lstwo}{\ensuremath{\small\text{\sf LS-2}}\xspace}

\newcommand{\lsj}{\ensuremath{\small\text{\sf LS-}j}\xspace}

\newcommand{\cmark}{\ding{51}\xspace}%
\newcommand{\xmark}{\ding{55}\xspace}%

\newcommand{\costfrac}{\ensuremath{\mathrm{cost}_{f}}\xspace}%
\newcommand{\POD}{\ensuremath{\text{\sc\large pod}}\xspace}%

\usepackage{graphicx}
\usepackage{xr}

\begin{document}
%

\title{Diversity-aware $k$-median:\\Clustering with fair center representation
\thanks{This research is supported by the Academy of Finland projects AIDA
(317085) and MLDB (325117), the ERC Advanced Grant REBOUND (834862), the EC
H2020 RIA project SoBigData (871042), and the Wallenberg AI, Autonomous Systems
and Software Program (WASP) funded by the Knut and Alice Wallenberg
Foundation.}}
\titlerunning{Diversity-aware $k$-median}

%
\author{Suhas Thejaswi\inst{1} \and Bruno Ordozgoiti\inst{1} \and Aristides
Gionis\inst{1,2}}
\authorrunning{Thejaswi, Ordozgoiti and Gionis}
%
\institute{Aalto University, Finland \\\email{firstname.lastname@aalto.fi} \and
KTH Royal Institute of Technology, Sweden \\\email{argioni@kth.se}}
\maketitle              
\begin{abstract}
We introduce a novel problem for diversity-aware clustering. We assume that the
potential cluster centers belong to a set of groups defined by protected
attributes, such as ethnicity, gender, etc. We then ask to find a minimum-cost
clustering of the data into $k$ clusters so that a specified minimum number of
cluster centers are chosen from each group. We thus require that all groups are
represented in the clustering solution as cluster centers, according to
specified requirements. 
More precisely, we are given a set of clients $C$, a set of facilities
$\pazocal{F}$, a collection $\mathcal{F}=\{F_1,\dots,F_t\}$ of facility groups
$F_i \subseteq \pazocal{F}$, budget $k$, and a set of lower-bound thresholds
$R=\{r_1,\dots,r_t\}$, one for each group in $\mathcal{F}$.  The
\emph{diversity-aware $k$-median problem} asks to find a set $S$ of $k$
facilities in $\pazocal{F}$ such that $|S \cap F_i| \geq r_i$, that is, at least
$r_i$ centers in $S$ are from group $F_i$, and the $k$-median cost 
$\sum_{c \in C} \min_{s \in S} d(c,s)$ is minimized.  
We show that in the general case where the facility groups may overlap, the diversity-aware $k$-median problem is \np-hard, fixed-parameter intractable, and inapproximable to any multiplicative factor.  On the other hand, when the facility groups are disjoint, approximation algorithms can be obtained by reduction to the \emph{matroid median} and \emph{red-blue median} problems.  Experimentally, we evaluate our approximation methods for the tractable cases, and present a relaxation-based heuristic for the theoretically intractable case, which can provide high-quality and efficient solutions for real-world datasets.

\end{abstract}

\cnote{Some comments (mainly stylistic):

--- ``Diversity-aware $k$-median'' sounds nicer than ``diversified $k$-median''.

--- For the set of facilities, can we use $F$ instead of $\pazocal{F}$?

--- For facility subsets $F_i$ perhaps we can use ``groups'' instead of ``types''?

}

\section{Introduction}

As many important decisions are being automated,
algorithmic fairness is becoming increasingly important.
Examples of critical decision-making systems include  
determining credit score for a consumer, 
computing risk factors for an insurance, 
pre-screening applicants for a job opening, 
dispatching patrols for predictive policing, and more. 
When using algorithms to make decisions for such critical tasks, 
it is essential to design and employ methods that minimize bias
and avoid discrimination
against people based on gender, race, or ethnicity.

Algorithmic fairness has gained wide-spread attention in the recent years~\cite{pedreshi2008discrimination}. 
The topic has been predominantly studied for \emph{supervised machine learning}, 
while fairness-aware formulations have also been proposed for \emph{unsupervised machine learning}, 
for example, 
\emph{fair clustering}~\cite{backurs2019scalable,bera2019fair,chierichetti2017fair,schmidt2019fair}, 
\emph{fair principal component analysis}~\cite{samadi2018price}, or 
\emph{fair densest-subgraph mining}~\cite{anagnostopoulos2020spectral}.
For the clustering problem the most common approach is to incorporate fairness 
by the means of \emph{representation-based constraints}, 
i.e., requiring that all clusters contain certain proportions of the different groups in the data, 
where data groups are defined via a set of protected attributes, such as demographics.
In this paper we introduce a novel notion for fair clustering
based on \emph{diversity constraints on the set of selected cluster centers}.

Research has revealed that bias can be introduced in
machine-learning algorithms when
bias is present in the input data used for training, and methods are
designed without considerations for diversity or constraints to
enforce fairness \cite{danks2017algorithmic}. 
A natural solution is to introduce diversity
constraints.
We can look at diversification from two different perspectives: 
($i$) avoiding over-representation; and 
($ii$) avoiding under-representation.
In this paper we focus on the latter requirement of avoiding under-representation.
Even though these two approaches look similar, 
a key contribution in our work is to observe that they are mathematically distinct
and lead to computational problems having different complexity; 
for details see Sections~\ref{sec:problem} and~\ref{section:tractable}.

To motivate our work, we present two application scenarios.

\spara{Committee selection:} 
We often select committees to represent an underlying population and work
towards a task, e.g., a program committee to select papers for a conference, 
or a parliamentary committee to handle an issue.
As we may require that each member of the population is represented by at least
one committee member, it is natural to formalize the committee-selection task
as a clustering problem, where the committee members will be determined by 
the centers of the clustering solution.
In addition, one may require that the committee is composed by a diverse mix
of the population with no underrepresented groups, e.g.,
a minimum fraction of the conference PC members work in industry, 
or a minimum fraction of the parliamentary committee are~women.

\spara{News-articles summarization:}
Consider the problem of summarizing a collection of news articles
obtained, for example, as a result to a user query.
Clustering these articles using a bag-of-words representation will allow us to select
a subset of news articles that cover the different topics present in the collection. 
In addition, one may like to ensure that the representative articles comes from a diverse
set of media sources, e.g., 
a minimum fraction of the articles comes from left-leaning media or from opinion columns.

\smallskip
To address the scenarios discussed in the previous two examples, 
we introduce a novel formulation of 
diversity-aware clustering with representation constraints on cluster centers.
In particular, we assume that a set of groups is associated with the facilities 
to be used as cluster centers. 
Facilities groups may correspond to demographic groups, in the first example, 
or to types of media sources, in the second. 
We then ask to cluster the data by selecting a subset of facilities as cluster centers, 
such that the clustering cost is minimized, 
and requiring that each facility group is \emph{not underrepresented} in the solution.

\begin{table}[t]
\footnotesize
\caption{\label{table:intro:overview}An overview of our results. 
All problem cases we consider are \np-hard.
{FPT($k$)} indicates whether the problem is
fixed-parameter tractable with respect to parameter $k$. 
\emph{Approx.\ factor} shows the factor of approximation obtained, 
and \emph{Approx.\ method} shows the method used.
}
\centering
\begin{tabular}{l c c c l}
\toprule
\multirow{2}{*}{Problem}   &  \multirow{2}{*}{~\np-hard~} &  \multirow{2}{*}{~FPT($k$)~} 
       &  ~~Approx.~~ & ~~Approx.~~\\
 &  &  &  ~~factor~~ & ~~method~~\\

\midrule
\multicolumn{5}{c}{Intractable case: intersecting facility groups} \\
\cmidrule(l{75pt}r{75pt}){1-5}
General variant & \cmark & \xmark &  \multicolumn{2}{l}{~~inapproximable} \\
\midrule
\multicolumn{5}{c}{Tractable cases: disjoint facility groups} \\
\cmidrule(l{75pt}r{75pt}){1-5}
$t > 2$, $\sum_{i \in [t]} r_i = k$ & \cmark & open & $8$ & ~~LP \\
$t > 2$, $\sum_{i \in [t]} r_i < k$ & \cmark & open & $8$ & ~~$\bigO(k^{t-1})$ calls to LP\\
$t = 2$, $r_1 + r_2 = k$ & \cmark &  open & $3+\epsilon$ & ~~local search \\
$t = 2$, $r_1 + r_2 < k$  & \cmark &  open & $3+\epsilon$ & ~~$\bigO(k)$ calls to local search \\
\bottomrule
\end{tabular}
\end{table}

We show that in the general case, where the facility groups overlap, 
the diversity-aware $k$-median problem is not only \np-hard, 
but also fixed-parameter intractable, 
and inapproximable to any multiplicative factor. In fact, we prove it
is \np-hard to even find a feasible solution, that is, a set of
centers which satisfies the representation constraints, regardless of
clustering cost.
These hardness results set our clustering problem in stark contrast with other clustering formulations
where approximation algorithms exist, and in particular, 
with the \emph{matroid-median problem}~\cite{Chen2016,Krishnaswamy2011,Hajiaghayi2010},
where one asks that facility groups are \emph{not over-represented}.
Unfortunately, however, the matroid-median problem does not ensure fairness for all facility groups.

On the positive side, 
we identify important cases for which the 
diversity-aware $k$-median problem is approximable, 
and we devise efficient algorithms with constant-factor approximation guarantees.
These more tractable cases involve settings when the facility groups are
disjoint. 
Even though the general variant of the problem in inapproximable, 
we demonstrate using experiments that we can obtain
a desired clustering solution with representation constraints with almost the same cost
as the unconstrained version using simple heuristics based on local-search.
%
%
The hardness and approximability results for the diversity-aware $k$-median problem 
are summarized in Table~\ref{table:intro:overview}.

In addition to our theoretically analysis and results
we empirically evaluate our methods on several real-world datasets. 
Our experiments show that in many problem instances, 
both theoretically tractable and intractable, 
\emph{the price of diversity is low in practice}.
In particular, our methods can be used to find solutions
over a wide range of diversity requirements
where the clustering cost is comparable to the cost of unconstrained clustering.

The rest of this paper is structured as
follows. Section~\ref{sec:related} discusses related work, 
Section~\ref{sec:problem} presents the problem statement and
computational complexity results. Section~\ref{section:tractable}
discusses special cases of the problem that admit polynomial-time
approximations. In Section~\ref{section:heuristics} we offer
efficient heuristics and related tractable objectives, and
in Section~\ref{section:experiments} we describe experimental results.
Finally, Section~\ref{section:conclusion} is a short conclusion.

\section{Related work}
\label{sec:related}
Algorithmic fairness has attracted a considerable amount of attention in recent years, 
as many decisions that affect us in everyday life are being made by algorithms.
Many machine-learning and data-mining problems have been adapted to incorporate notions of fairness. 
Examples include problems in 
classification~\cite{barocas-hardt-narayanan,dwork2012fairness,hardt2016equality}, 
ranking~\cite{singh2019policy,zehlike2017fa}, 
recommendation systems~\cite{yao2017beyond}, 
and more. 

In this paper we focus in the problem of clustering
and we consider a novel notion of fairness based on diverse representation of cluster centers. 
Our approach is significantly different (and orthogonal) from the 
standard notion of \emph{fair clustering}, 
introduced by the pioneering work of Chierichetti et al.~\cite{chierichetti2017fair}.
In that setting, data points are partitioned into groups (Chierichetti et al.\ considered only two groups)
and the fairness constraint is that each cluster should a certain fraction of points from each group.
Several recent papers have extended the work of Chierichetti et al.\
by proposing more scalable algorithms~\cite{backurs2019scalable,huang2019coresets}, 
extending the methods to accommodate more than two groups~\cite{bercea2019cost,schmidt2019fair}, 
or introducing privacy-preserving properties~\cite{rosner2018privacy}.
In this line of work, the fairness notion applies to the 
representation of data groups within each cluster. 
In contrast, in our paper the fairness notion applies 
to the representation of groups in the cluster centers.

The closest-related work to our setting are the problems of 
\emph{matroid median}~\cite{Chen2016,Krishnaswamy2011}
and \emph{red-blue median}~\cite{Hajiaghayi2010,hajiaghayi2012local},
which can be used to ensure that no data groups are over-represented 
in the cluster centers of the solution ---
in contrast we require that no data groups are underrepresented.
Although the two notions are related, in a way that we make precise in Section~\ref{section:tractable},
in the general case they differ significantly, 
and they yield problems of greatly different complexity.
Furthermore, in the general case, the matroid-median problem cannot be used
to ensure fair representation, 
as upper-bound constraints cannot ensure 
that all groups will be represented in the solution.
Although it is for those cases that our diversity-aware clustering problem is intractable,
one can develop practical heuristics that achieve fair results, 
with respect to diverse representation,  
as shown in our experimental evaluation.

\section{Problem statement and complexity}
\label{sec:problem}

We consider a set of clients $C$ and a set of facilities $\pazocal{F}$. 
In some cases, the set of facilities may coincide with the set of clients ($\pazocal{F}=C$), 
or it is a subset  ($\pazocal{F}\subseteq C$). 
We assume a distance function $d:C \times \pazocal{F} \rightarrow \realsnn$, 
which maps client--facility pairs into nonnegative real values.
We also consider a collection $\mathcal{F}=\{F_1,\dots,F_t\}$ of facility groups $F_i \subseteq \pazocal{F}$.
During our discussion we distinguish different cases for the structure of $\mathcal{F}$. 
In the most general case 
the facility groups $F_i$ may overlap. 
Two special cases of interest, discussed in Section~\ref{section:tractable}, 
are when the facility groups $F_i$ are disjoint and when there are only two groups. 
Finally,  we are given a total budget $k$ of facilities to be selected,
and a set of lower-bound thresholds $R=\{r_1,\dots,r_t\}$, 
i.e., one threshold for each group $F_i$ in $\mathcal{F}$. 

The {diversity-aware $k$-median problem} (\divkmedian)
asks for a set $S$ of $k$ facilities in $\pazocal{F}$
subject to the constraint $|S \cap F_i| \geq r_i$, such that the
$k$-median cost $\sum_{c \in C} \min_{s \in S} d(c,s)$ is minimized.  
Thus, we search for a minimum-cost clustering solution $S$
where each group $F_i$ is represented by at least $r_i$ centers.

In the following sections, we study the computational complexity of the \divkmedian
problem. In particular, we show that the general variant of the problem is
($i$) \np-hard; 
($ii$) not fixed-parameter tractable with respect to parameter $k$,
i.e., the size of the solution sought; and 
($iii$) inapproximable to any multiplicative factor.
%
In fact, we show that hardness results ($i$) and ($ii$) apply for the problem of simply finding a
feasible solution. That is, in the general case, and assuming $\p \neq \np$
there is no polynomial-time algorithm to find a solution 
$S \subseteq \pazocal{F}$ that satisfies the constraints $|S \cap F_i| \geq r_i$, for all $i \in [t]$.
The inapproximability statement ($iii$) is a consequence of the \np-hardness 
for finding a feasible solution.
These hardness results motivate the heuristics we propose later on.

\subsection{\np-hardness}
We prove \np-hardness by reducing the dominating set
problem to the problem of finding a {\em feasible solution} to \divkmedian.


\spara{Dominating set problem (\dominatingset).} Given a graph $G=(V, E)$ with $|V|=n$ vertices, 
and an integer $k \leq n$,
decide if there exists a subset $S \subseteq V$ of size $|S|=k$
such that for each $v \in V$ it is either $\{v\} \cap S \neq \emptyset$ or 
$\{v\} \cap N(S) \neq \emptyset$, 
where $N(S)$ denotes the set of vertices adjacent to at least one vertex in $S$. 
In other words, each vertex in $V$ is either in $S$ or
adjacent to at least one vertex in $S$.

\begin{lemma}
\label{lemma:hardness:nphard}
Finding a feasible solution for \divkmedian is \np-hard.
\end{lemma}
\begin{proof}
Given an instance of \dominatingset $(G=(V,E), k)$, we
construct an instance of the \divkmedian problem $(C, \pazocal{F}, \mathcal{F}, d, k, R)$, 
such that $C=V$, $\pazocal{F}=V$, $d(u,v)=1$ for all $(u,v) \in C \times F$, 
$\mathcal{F} = \{F_1,\dots,F_n\}$ with $F_u = \{u\} \cup N(u)$,
and $R=\{1,\ldots,1\}$, 
i.e., the lower-bound thresholds are set to $r_u=1$, for all $u\in V$.

Let $S \subseteq C$ be a feasible solution for \divkmedian. 
From the construction it is clear that $S$ is a dominating set, as $|F_u \cap S| \geq 1$, 
and thus $S$ intersects $\{u\} \cup N(u)$ for all $u\in V$.
The proof that a dominating set is a feasible solution to \divkmedian is analogous.
\qed
\end{proof}
The hardness of diversity-aware $k$-median follows immediately.
\begin{corollary}
\label{corollary:hardness:nphard}
The \divkmedian problem is \np-hard.
\end{corollary}

\subsection{Fixed-parameter intractability}

A problem $P$ specified by input $x$ and a parameter $k$ 
is {\em fixed-parameter tractable} (FPT) if there exists an algorithm
$A$ to solve every instance $(x,k) \in P$ with running time of the form 
$f(k) |x|^{\bigO(1)}$, where $f(k)$ is function depending solely on the parameter $k$ and
$|x|^{\bigO(1)} = \mathrm{poly}(|x|)$ is a polynomial independent of the parameter $k$. 
A problem $P$ is {\em fixed-parameter intractable} otherwise

%

To show that the \divkmedian is fixed-parameter intractable we present a
{\em parameterized reduction} from the \dominatingset problem to \divkmedian.%
\footnote{Let $P$ and $P'$ be two problems. 
A {\em parameterized reduction} from $P$
to $P'$ is an algorithm $A$ that transforms an instance $(x, k) \in P$ to an
instance $(x',k') \in P'$ such that: 
($i$) $(x,k)$ is a {\sf yes} instance of $P$ if and only if $(x',k')$ is a {\sf yes} instance of~$P'$; 
($ii$) $k' \leq g(k)$ for some computable function $g$; and 
($iii$) the running time of the transformation $A$ is $f(k)|x|^{\bigO(1)}$, for some computable function $f$. 
Note that $f$ and $g$ need not be polynomial functions. 
For details see Cygan et al.~\cite[Chapter~13]{cygan2015parameterized}.}
%
%
The \dominatingset problem is known to be fixed-parameter intractable
\cite[Theorem~13.9]{cygan2015parameterized}. This means that there exists no
algorithm with running time $f(k)|V|^{\bigO(1)}$ to solve \dominatingset, 
where $f(k)$ is a function depending solely on the parameter $k$.
%

\begin{theorem}
\label{theorem:hardness:fpt}
The \divkmedian problem is fixed-parameter intractable with respect to the parameter
$k$, that is, the size of the solution sought.
\end{theorem}
\begin{proof}
We apply the reduction from Lemma~\ref{lemma:hardness:nphard}.
It follows that 
($i$) an instance $(G,k)$ of the \dominatingset problem has a feasible
solution if and only if there exists a feasible solution for the \divkmedian
problem instance $(C, \pazocal{F}, \mathcal{F}, d, k', R)$, with $k'= k$, and 
($ii$) the reduction takes polynomial time in the size of the input. 
So there exists a
parameterized reduction from the \dominatingset problem to the \divkmedian problem.
This implies that if there exists an algorithm with running time 
$f(k')|C|^{\bigO(1)}$ for the \divkmedian problem then there exits an algorithm
with running time $f(k)|V|^{\bigO(1)}$ for solving the \dominatingset problem.  
\qed
\end{proof}

It would still be interesting to
check whether there exists a parameter of the problem that can be used to design a
solution where the exponential complexity can be restricted. We leave this as an open problem.



\subsection{Hardness of approximation}

We now present hardness-of-approximation results for \divkmedian. 

\begin{theorem}
\label{theorem:hardness:approx}
Assuming $\p \neq \np$, the \divkmedian problem cannot be approximated to any
multiplicative factor.
\end{theorem}
\begin{proof}
We apply the reduction of the \dominatingset problem
from the proof of Lemma~\ref{lemma:hardness:nphard}. 
For the sake of contradiction let $A$ be a polynomial-time approximation algorithm, 
which can be used to obtain a factor-$c$ approximate solution for \divkmedian.
Then we can employ algorithm $A$ to obtain an exact solution to the
\dominatingset instance in polynomial time, by way of the aforementioned reduction. 
The reason is that an approximate solution for \divkmedian is also a feasible solution, 
which in turn implies a feasible solution for \dominatingset.
Thus, unless $\p \neq \np$, \divkmedian cannot be approximated to any multiplicative factor.
\qed
\end{proof}

We observe that this inapproximability result applies even to certain special cases of the problem, 
where the input has special structure.
The proofs of Theorem~\ref{theorem:hardness:approx2} and
Theorem~\ref{theorem:hardness:approx3} are available in the Supplementary
Section~\ref{sup:proofs}.

\begin{theorem}
\label{theorem:hardness:approx2}
Assuming $\p \neq \np$, the \divkmedian problem cannot be approximated to any
multiplicative factor even if all the subsets in $\mathcal{F}$ have size~$2$.
\end{theorem}


\begin{theorem}
\label{theorem:hardness:approx3}
Assuming $\p \neq \np$, the \divkmedian problem cannot be approximated to any
multiplicative factor even if the underlying metric space is a tree metric.
\end{theorem}

\section{Approximable instances}
\label{section:tractable}

In the previous section we presented strong intractability results for the \divkmedian problem.
Recall that inapproximability stems from the fact that satisfying the
non under-representation constraints $|S \cap F_i| \geq r_i$ for $i \in [t]$ is \np-hard.
However, there are instances where finding a feasible solution is
polynomial-time solvable, even if finding an minimum-cost clustering solution remains \np-hard.
%
%
In this section we discuss such instances and give approximation algorithms.

\subsection{Non-intersecting facility groups}

We consider instances of \divkmedian where
$F_i \cap F_j = \emptyset$ for all $F_i, F_j \in \mathcal{F}$,
that is, the facility groups are disjoint. 
We refer to variants satisfying disjointness conditions
as the \divkmediannif problem.

It is easy to verify that a feasible solution exists for \divkmediannif
if and only if $|F_i|\ge r_i$ for all $i\in[t]$ and $\sum_{i \in [t]} r_i \leq k$.
Furthermore, assuming that the two latter conditions hold true, 
finding a feasible solution is trivial:
it can be done simply by picking $r_i$ facilities from each facility group $F_i$.

It can be shown that the \divkmediannif problem can be reduced to the 
\emph{matroid-median problem}~\cite{Krishnaswamy2011}, 
and use existing techniques for the latter problem to 
obtain an $8$-approximation algorithm for \divkmediannif~\cite{swamy2016improved}.
Before discussing the reduction we first introduce the matroid-median problem.

\spara{The matroid-median problem} (\matmedian)~\cite{Krishnaswamy2011}.
We are given a finite set of clients $C$ and facilities $\pazocal{F}$, 
a metric distance function $d: C \times \pazocal{F} \rightarrow \realsnn$, 
and a matroid $\pazocal{M}=(\pazocal{F},\pazocal{I})$ with ground set $\pazocal{F}$
and a collection of independent sets $\pazocal{I} \subseteq 2^\pazocal{F}$. 
The problem
asks us to find a subset $S \in \pazocal{I(M)}$ such that the cost function
$\mathrm{cost}(S)=\sum_{c \in C} \min_{s \in S} d(c,s)$ is minimized.
\smallskip

The \matmedian problem is a generalization of $k$-median, and has an $8$-approximation algorithm based on
LP relaxation~\cite{swamy2016improved}. Here we present a reduction of
 \divkmediannif to \matmedian. 
In this section we handle the case where $\sum_{i \in [t]} r_i = k$. 
In Section \ref{section:sumr_i<k} we show that the case $\sum_{i \in [t]} r_i < k$ 
can be reduced to the former one with at most $\bigO(k^{t-1})$ calls. 
Approximating \divkmediannif in polynomial-time when $\sum_{i \in [t]} r_i < k$ is left open.

\spara{The reduction.} Given an instance $(C, \pazocal{F}, \mathcal{F}, d, k, R)$, 
of the \divkmediannif problem we generate an instance $(C',\pazocal{F'},\pazocal{I'},d')$ 
of the \matmedian problem as follows: 
$C'= C$, $\pazocal{F'}=\pazocal{F}$, $d'=d$,
and $\pc{M}=(\pc{F'},\pc{I'})$ where $\pc{I'} \subseteq 2^\pc{F'}$
and $A \in \pc{I'}$ if $|A \cap F_i| \leq r_i$ for all $r_i \in R$. 
More precisely, the set of independent sets is comprised of all subsets of $\pc{F'}$ 
that satisfy \emph{non over-representation} constraints. 
It is easy to verify that $\pc{M}$ is a matroid --- it is a \emph{partition matroid}.
In the event that the algorithm for \matmedian outputs a solution 
where $|A \cap F_i| < r_i$, for some $i$, 
since $\sum_{i \in [t]} r_i = k$, it is trivial to satisfy all the constraints with equality by
completing the solution with facilities of the missing group(s) at no additional connection cost.
Since we can ensure that $|A \cap F_i| = r_i$, for all $i$, 
it also holds $|A \cap F_i| \ge r_i$, for all $i$, 
that is, the \divkmediannif constraints.

Since the \matmedian problem has a polynomial-time approximation
algorithm, it follows from our inapproximability results (Section~\ref{sec:problem}) that a reduction of the general formulation of
\divkmedian is impossible.
We can thus conclude that allowing intersections between 
facility groups fundamentally changes the combinatorial structure of 
feasible solutions, interfering with the design of approximation
algorithms. 

\subsection{Two facility groups}
\label{section:rb-median}

The approximation guarantee of the \divkmediannif problem
can be further improved if we restrict the number of groups to two. 

In particular, we consider instances of the \divkmedian problem where
$F_i \cap F_j = \emptyset$, for all $F_i, F_j \in \mathcal{F}$, and $\mathcal{F}=\{F_1,F_2\}$.
For simplicity, the
facilities $F_1$ and $F_2$ are referred to as \emph{red} and \emph{blue} facilities, respectively.

As before, we can assume that $\sum_{i \in [t]} r_i = r_1+r_2\leq k$, 
otherwise the problem has no feasible solution.
We first present a local-search algorithm for the case $r_1 + r_2 = k$. 
In Section \ref{section:sumr_i<k} we show that the case with $r_1 + r_2 < k$ 
can be reduced to the former one with a linear number of calls for 
different values of $r_1$ and $r_2$. 
Before continuing with the algorithm we first define the \rbmedian problem.


\mpara{The red-blue median problem (\rbmedian).} 
We are given a set of clients~$C$,
two disjoint facility sets  $F_1$ and $F_2$
(referred to as red and blue facilities, respectively), 
two integers $r_1, r_2$ and a metric distance function
$d: C \times \{F_1 \cup F_2\} \rightarrow\realsnn$. 
The problem asks to find a subset $S \subseteq F_1 \cup F_2$ such that 
$|F_1 \cap S| \leq r_1$, $|F_2 \cap S| \leq r_2$ and the cost function
$\mathrm{cost}(S) = \sum_{c \in C} \min_{s \in S} d(c,s)$ is minimized. 
\smallskip

The \rbmedian problem accepts a $3 + \epsilon$ approximation algorithm based on
local-search~\cite{hajiaghayi2012local}. The algorithm works by swapping a
red-blue pair $(r, b)$ with a red-blue pair $(r', b')$ as long as the
cost improves. 
Note that $(r' = r, b' \neq b)$, $(r' \neq r, b'=b)$ and 
$(r' \neq r, b' \neq b)$ are valid swap pairs.
The reduction of \divkmedian to \rbmedian is similar to the one given above for
\matmedian. Thus, when the input consists of two non-intersecting
facility groups we can obtain a $3 + \epsilon$ approximation of the
optimum in polynomial time which follows from the local-search approximation of the
\rbmedian problem.

\subsection{The case $\sum_ir_i<k$}
\label{section:sumr_i<k}

The reduction of \divkmediannif to \matmedian relies on picking exactly $\sum_i r_i$ facilities. 
This is because it is not possible to define a matroid that simultaneously enforces the desired 
lower-bound facility group constraints and the cardinality constraint for the solution. 
Nevertheless, we can overcome this obstacle at a bounded cost in running time.

So, in the case that $\sum_ir_i<k$, 
in order to satisfy the constraint $|S|=k$, we can simply increase the
lower-bound group constraints $r_i\mapsto r_i'>r_i$, 
$i=1, \dots, t$ so that $\sum_i r_i'=k$. 
However, if we do this in an arbitrary fashion we might make a suboptimal choice.
To circumvent this, we can exhaustively inspect all possible choices. 
For this, it suffices to construct ${k-\sum_ir_i+t-1 \choose t-1}=\bigO(k^{t-1})$ 
instances of \matmedian. 
In the case of \rbmedian discussed in Section~\ref{section:rb-median}, 
i.e., when $r_1+r_2<k$, the required number of instances is linear in~$k$.

\cnote[Aris]{Why quadratic? 
Isn't it exactly $k-(r_1+r_2)+1$?}

\section{Proposed methods}
\label{section:heuristics}

In this section we present practical methods to solve the diversity-aware clustering problem.
In particular, we present local-search algorithms for \divkmediannif
and a method based on relaxing the representation constraints for \divkmedian.



\subsection{Local search}

Algorithms based on the local-search heuristic have been used to design
approximation algorithms for many optimization problems, including facility
location~\cite{Arya2012,charikar1999improved,gupta2008simpler} and
$k$-median~\cite{Arya2012,charikar1999improved,hajiaghayi2012local} problems.
In light of the inapproximability results presented in the previous section it
comes as no surprise that any polynomial-time algorithm, including local-search
methods, cannot be expected to find a feasible solution for the \divkmedian
problem. Nevertheless, local-search methods are viable for the tractable
instances discussed in Section~\ref{section:tractable}, and can be shown to
provide provable quality guarantees.

For solving the \divkmediannif problem we propose two algorithms based
on local search.

\spara{Local search variant\,\#1 (\lsone).}
We propose a single-swap local-search algorithm 
described in Figure~\ref{figure:localsearch}. The key difference with respect to
vanilla local search is that we must ensure that a
swap does not violate the representation constraints.

\begin{figure}[t]
\begin{enumerate}
\item Initialize $S$ to be an arbitrary feasible solution.
\item While there exists a pair $(s,s')$, with $s \in S$ and $s' \in \pc{F}$ such that
  \begin{enumerate}
  \item $\mathrm{cost}(S \setminus \{s\} \cup \{s'\}) < \mathrm{cost}(S)$ and 
  \item $S \setminus \{s\} \cup \{s'\}$ is feasible i.e, $|S \setminus \{s\}
\cup \{s'\} \cap F_i| \geq r_i$ for all $i \in [t]$,
  \end{enumerate}
   Set $S = S \setminus \{s\} \cup \{s'\}$.
\item Return $S$.
\end{enumerate}
\caption{Local search heuristic (\lsone) for \divkmediannif.}
\label{figure:localsearch}
\end{figure}

\begin{figure}[t]
\centering
\begin{tikzpicture}[baseline=-1.25em,
    roundnode/.style={circle, draw=gray!60, fill=gray!20, very thick, minimum size=7mm},
  ]
  \node[roundnode, diagonal fill={red!70}{blue!80}, minimum size=.5cm] at (0,2)  (a1)                              {$f_1$};
  \node[roundnode, minimum size=.5cm, right = 1.3cm  of a1]   (a2) {};
  \node[roundnode, diagonal fill={red!70}{yellow!80}, minimum size=.5cm, right = 1.3cm of a2]  (a3)  {$f_3$};
  \node[roundnode, minimum size=.5cm, below = 1.3cm of a2]     (a4)  {};
  \node[roundnode, diagonal fill={green!70}{yellow!80}, minimum size=.5cm, left = 1.3cm of a4]     (a5)  {$f_2$};
  \node[roundnode, diagonal fill={green!80}{blue!80}, minimum size=.5cm, right = 1.3cm of a4]     (a6)  {$f_4$};
  
  \draw (a1) -- (a2) node[above, pos=0.5, black]{c};
  \draw (a1) -- (a4) node[below, pos=0.6, black]{c};
  \draw (a3) -- (a2) node[above, pos=0.5, black]{1};
  \draw (a3) -- (a4) node[below, pos=0.6, black]{1};
  \draw (a5) -- (a2) node[above, pos=0.6, black]{c};
  \draw (a5) -- (a4) node[below, pos=0.5, black]{c};
  \draw (a6) -- (a2) node[above, pos=0.6, black]{1};
  \draw (a6) -- (a4) node[below, pos=0.5, black]{1};
\end{tikzpicture}
\caption{An example illustrating the infeasibility of local search.}
\label{fig:localsearch:1}
\end{figure}
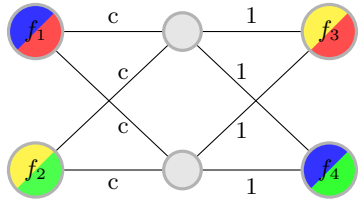

We stress that the proposed algorithm \lsone is not viable for
general instances of \divkmedian with intersecting facility groups. To
illustrate, we present an
example in Figure~\ref{fig:localsearch:1}.
Let $F_r,F_g,F_b,F_y$ be facility groups, corresponding to the
colors red, green, blue and yellow, respectively. The intersection
cardinality constraints $r_r=r_g=r_b=r_y=1$ and the number of medians $k=2$.

Let $S=\{f_1,f_2\}$ be a feasible solution. 
It is trivial to see that we cannot swap $f_1$ with either $f_3$ or $f_4$, 
since both swaps violate the constraints $|S \cap F_b| \geq 1$ and $|S \cap F_r| \geq 1$, respectively. 
Likewise we cannot swap $f_2$ with either $f_3$ or $f_4$. 
So our local-search algorithm is stuck at a local optima
and the approximation ratio is $c$, which can be made arbitrarily large.
We can construct a family of infinitely many such problem instances where the 
local-search algorithm returns arbitrarily bad results.
Similarly we can construct a family of infinitely many instances where the
\divkmedian problem with $t$ facility groups and $k<t$ would require at least
$t-1$ parallel swaps to ensure that local search is not stuck in a local optima. 
This example illustrates the limited applicability of the local-search heuristic for the 
most general variant of the \divkmedian problem, where the facility groups 
overlap in an arbitrary way.

%


\spara{Local search variant\,\#2 (\lstwo).} 
Our second approach is the multi-swap
local-search heuristic described in Figure~\ref{fig:localsearch:2}. The algorithm works by
picking $r_i$ facilities from $F_i$ and $k - \sum_{i \in [t]} r_i$ from
$\pc{F}$ as an initial feasible solution. 
We swap a tuple of facilities $(s_1,\dots,s_{t+1})$ with
$(s'_1,\dots,s'_{t+1})$ as long as the
cost improves. 
The algorithm has running time of $\bigO(n^t)$, and thus
it is not practical for large values of $t$. 
The algorithm \lstwo is a $3+\epsilon$ approximation for the \divkmediannif problem
with two facility groups i.e., $t=2$ (see Section~\ref{section:rb-median}).
Bounding the approximation ratio of algorithm \lstwo for $t > 2$ 
is an open problem.

\begin{figure}[t]
\begin{enumerate}
\item Initialize --- arbitrarily pick:
\begin{enumerate}
  \item $S_i \subseteq F_i$ such that $|S_i| = r_i$ for all $i \in [t]$,
  \item $S_{t+1} \subseteq \pc{F} \setminus \bigcup_{i \in [t]} S_i$ such that 
        $|S_{t+1}| = k - \sum_{i \in [t]} r_i$, and
  \item initial solution is $S = \bigcup_{i \in [t]} S_i \cup S_{t+1}$.
\end{enumerate}
\item Iterate --- while there exists tuples $(s_1,\dots,s_{t+1})$ and 
$(s'_1,\dots,s'_{t+1})$ such that:
\begin{enumerate}
  \item $s_i \in S_i$, $s'_i \in F_i$ for all $i \in [t]$, $s_{t+1} \in S_{t+1}$, \
        $s'_{t+1} \in \pc{F} \setminus \bigcup_{i \in [t]} S_i$
  \item $S \setminus \{s_1,\dots,s_{t+1}\} \cup \{s'_1,\dots,s'_{t+1}\}$ is
feasible, and
  \item $\mathrm{cost}(S \setminus \{s_1,\dots,s_{t+1}\} \cup \{s'_1,\dots,s'_{t+1}\}) < \mathrm{cost}(S)$
\end{enumerate}
  set $S = S \setminus \{s_1,\dots,s_{t+1}\} \cup \{s'_1,\dots,s'_{t+1}\}$.
\item Return $S$.
\end{enumerate}
\caption{Local-search heuristic (\lstwo) for \divkmediannif}
\label{fig:localsearch:2}
\end{figure}

\subsection{Relaxing the representation constraints}

In view of the difficulty of solving the problem as formulated in Section~\ref{sec:problem}, we explore
alternative, more easily optimized formulations to encode the
desired representation constraints.
We first observe that a straightforward approach,
akin to a Lagrangian relaxation, might result in undesirable outcomes. 
Consider the following
objective function:
\begin{equation}
\mathrm{cost}(S) = \sum_{v \in C} \min_{s \in S} d(v,s) + 
          \lambda \sum_{i \in [t]} \max\{r_i - |F_i\cap S|, 0\} , 
\end{equation}
that is, instead of enforcing the constraints, we penalize their violations. 
A problem with this formulation is that every constraint
satisfaction --- up to $r_i$ --- counts the same, and thus
the composition of the solution might be imbalanced. 

We illustrate this shortcoming with a simple example.
Consider ${\mathcal F}=\{F_1, F_2, F_3\}$, $k=6$, $r_1=2$, $r_2=2$, $r_3=0$. 
Now consider two solutions:
($i$)
$2$ facilities from $F_1$, $0$ from $F_2$, and $4$ from $F_3$;
and ($ii$)
$1$ facility from $F_1$, $1$ from $F_2$, and $4$ from $F_3$.
Both solutions score the same in terms of number of violations. 
Nevertheless, the second one is more balanced in terms of
group representation. To overcome this issue, we propose the following
alternative formulation.
\begin{equation}
\label{eq:relaxed-cost:2}
\costfrac(S) = \sum_{v \in C} \min_{s \in S} d(v,s) + 
          \lambda \sum_{i \in [t]} \frac{r_i}{|S \cap F_i| + 1}.
\end{equation}
The second term that encodes the violations enjoys group-level diminishing returns. 
Thus, when a facility of a protected group is added,
facilities from other groups will be favored. The cardinality
requirements $r_i$ act here as weights on the different groups.

We optimize the objective in Equation~\ref{eq:relaxed-cost:2} using vanilla
local-search by picking an arbitrary initial solution with no restrictions.

\section{Experimental evaluation}
\label{section:experiments}

In order to gain insight on the proposed problem and to evaluate our
algorithms, we carried out experiments on a variety of publicly
available datasets. 
Our objective is to evaluate the following key aspects:

\smallskip
\noindent
\textit{Price of diversity:} What is the price of enforcing
  representation constraints? We measure how the clustering cost
  increases as more stringent requirements on group representation are imposed.

\smallskip
\noindent
\textit{Relaxed objective:} 
  We evaluate the relaxation-based method, described in Section~\ref{section:heuristics}, 
  for the intractable case with facility group intersections. 
  We evaluate its performance in
  terms of constraint satisfaction and clustering cost.

\smallskip
\noindent
\textit{Running time:} Our problem formulation requires
  modified versions of standard local-search heuristics, 
  as described in Section~\ref{section:heuristics}. We evaluate
  the impact of these variants on running time.

\spara{Datasets.}
We use datasets from the UCI machine learning
repository~\cite{Dua:2019}. We normalize columns to unit norm and use 
the $L_1$ metric 
as distance function. 
The dataset statistics are reported in Table~\ref{table:datasets}. 

\begin{table}[t]
\centering
\caption{\label{table:datasets}
Dataset statistics. $n$ is the number of data points, $D$ is
dataset dimension, $t$ is number of facility types. 
Columns 4,5 and 6, 7 is the maximum and minimum size of
facility groups when divided into two disjoint groups and four intersecting
groups, respectively.}
\footnotesize
\begin{tabular}{l r r r r r r }
\toprule
 & & & \multicolumn{2}{c}{$t=2$} & \multicolumn{2}{c}{$t=4$}\\
          Dataset &  ~~$n$~~ & ~~$D$~~ & ~~min~~ & ~~max~~ & ~~min~~ & ~~max~~ \\
\midrule
 heart-switzerland &     123 & 14 &     10 &    113 &      - &      -\\
          heart-va &     200 & 14 &      6 &    194 &      - &      -\\
   heart-hungarian &     294 & 14 &     81 &    213 &      - &      -\\
   heart-cleveland &     303 & 14 &     97 &    206 &      - &      -\\
       student-mat &     395 & 33 &    208 &    187 &      - &      -\\
    house-votes-84 &     435 & 17 &    267 &    168 &      - &      -\\
       student-por &     649 & 33 &    383 &    266 &      - &      -\\
      student-per2 &     666 & 12 &    311 &    355 &      - &     - \\
            autism &     704 & 21 &    337 &    367 &     20 &    337\\
      hcv-egy-data &  1\,385 & 29 &    678 &    707 &     40 &    698\\
               cmc &  1\,473 & 10 &    220 & 1\,253 &     96 &    511\\
           abalone &  4\,177 &  9 & 1\,307 & 1\,342 &      - &     - \\
          mushroom &  8\,123 & 23 & 3\,375 & 4\,748 & 1\,072 & 3\,655\\
          nursery  & 12\,959 &  9 & 6\,479 & 6\,480 & 3\,239 & 4\,320\\
\bottomrule
\end{tabular}
\end{table}


\spara{Baseline.}
%
%
%
As a baseline we use a local-search algorithm with no cardinality constraints. 
We call this baseline \lszero. 
For each dataset we perform $10$ executions of \lszero with random initial assignments to
obtain the solution with minimum cost $\ell_0$ among the independent executions.  
\lszero is known to provide a $5$-approximation for the $k$-median problem~\cite{Arya2012}. 
We also experimented with {\em exhaustive enumeration} and {\em linear program solvers}, 
however these approaches failed to solve instances with modest size, 
which is expected given the the inherent complexity of \divkmedian. 
For details see Supplementary Section~\ref{sup:baseline}.

\spara{Experimental setup.}
The experiments are executed on a desktop with 4-core {\em Haswell}
CPU and $16$\,GB main memory. Our source code is written in {\tt Python}
and we make use of {\tt numpy} to enable parallelization of computations. 
Our source code is
anonymously available as open source~\cite{ourcode}.

\subsection{Results}

\spara{Price of diversity.}
For each dataset 
we identify a protected attribute and classify data points into two
disjoint groups. In most datasets we choose gender, except
in {\tt house-votes} dataset where we use party affiliation. We identify the
smallest group in the dataset ({\em minority group}) and measure the
fraction of the chosen facilities that belong to that group ({\em minority fraction}).
When running \lsone and \lstwo we enforce a specific minority fraction and repeat the
experiments for ten iterations by choosing random initial 
assignments. We refer to the cost of the solutions obtained from \lszero, \lsone and
\lstwo as $\ell_0$, $\ell_1$ and $\ell_2$, respectively.

The {\em price of diversity} (\POD) is the ratio of increase in the cost of the
solution to the cost of $LS_0$ i.e., 
$\POD(\lsone) = \frac{\ell_1 - \ell_0}{\ell_0}$
and $\POD(\lstwo) = \frac{\ell_2 - \ell_0}{\ell_0}$. 
Recall that in theory the \POD is unbounded.
However, this need not be the case in  practical scenarios.
Additionally, we compute the differences in group representation
between algorithms as follows. Let $R^i=\{r^i_1,\dots,r^i_t\}$ be the set
representing the number of facilities chosen from each group in $\mathcal{F}$ by
algorithm \lsj.
For $j=1,2$ we define $L_1(\lsj) = \sum_{i \in [t]} |r^j_i-r^0_i|/(k t)$. 

\begin{figure}[t]
\includegraphics[width=1.0\linewidth]{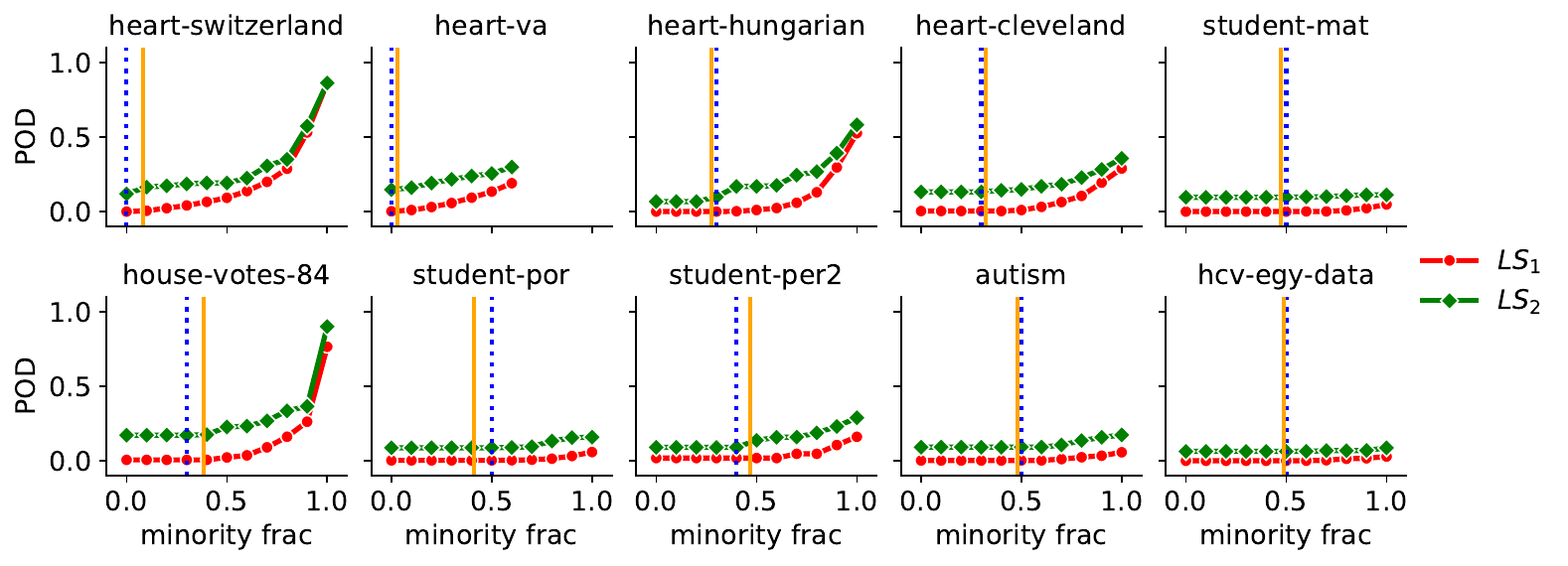}
\caption{\label{fig:experiments:pod}
Price of diversity ($k=10$). }
\end{figure}

In Figure~\ref{fig:experiments:pod}, we report the price of diversity (\POD) as a function of the
imposed minority fraction for \lsone and \lstwo. 
The blue and yellow vertical lines denote the minority
fraction achieved by the baseline \lszero and the fraction of minority
facilities in the dataset, respectively.
Notice that the minority fraction of the baseline is very close to the
minority fraction of the dataset.
With respect to our methods \lsone and \lstwo, 
we observe little variance 
among the independent executions. 
Most importantly, we observe that the price of diversity is relatively low, 
that is, for most datasets we can vary the representation requirements 
over a wide range and the clustering cost increases very little compared to the non-constrained version.
An increase is observed only for a few datasets and only for extreme values 
of representation constraints.
We also observe that \lsone outperforms consistently \lstwo.
This is good news as \lsone is also more efficient.


\begin{figure}[t]
\includegraphics[width=1.0\linewidth]{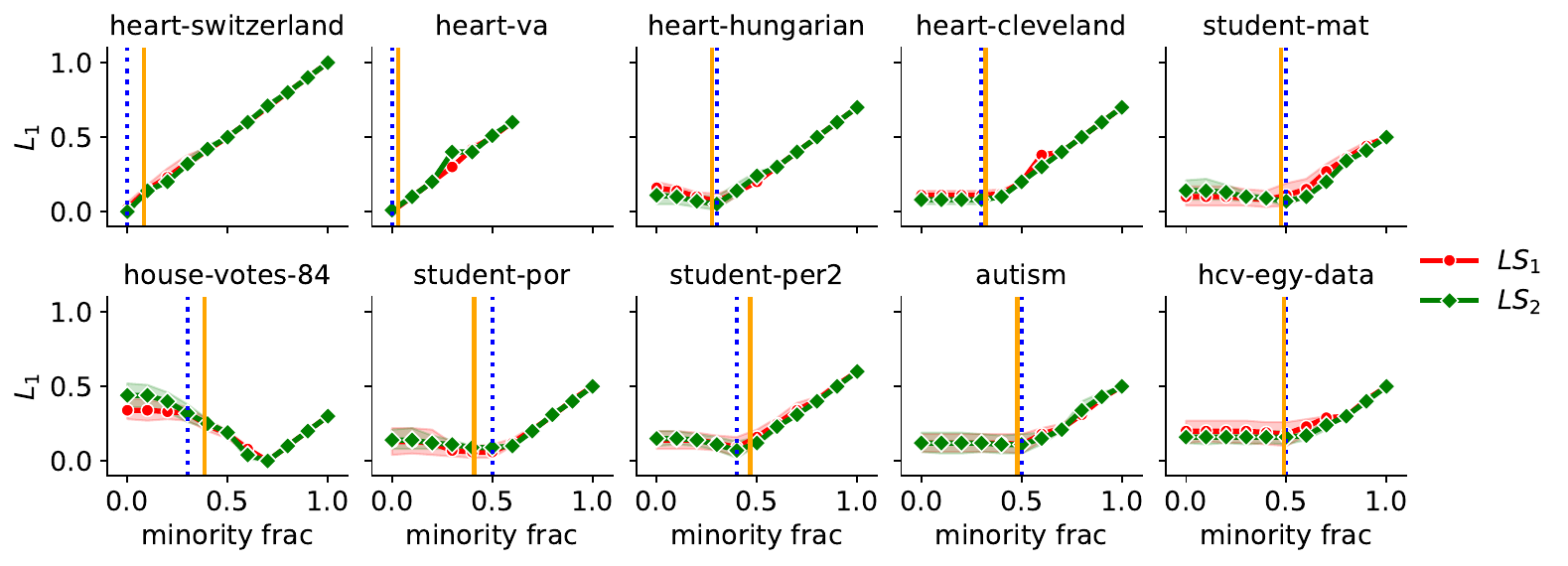}
\caption{\label{fig:experiments:l1}
$L_1$ distance of the chosen facility types ($k=10$). }
\end{figure}

\begin{figure}[t]
\includegraphics[width=1.0\linewidth]{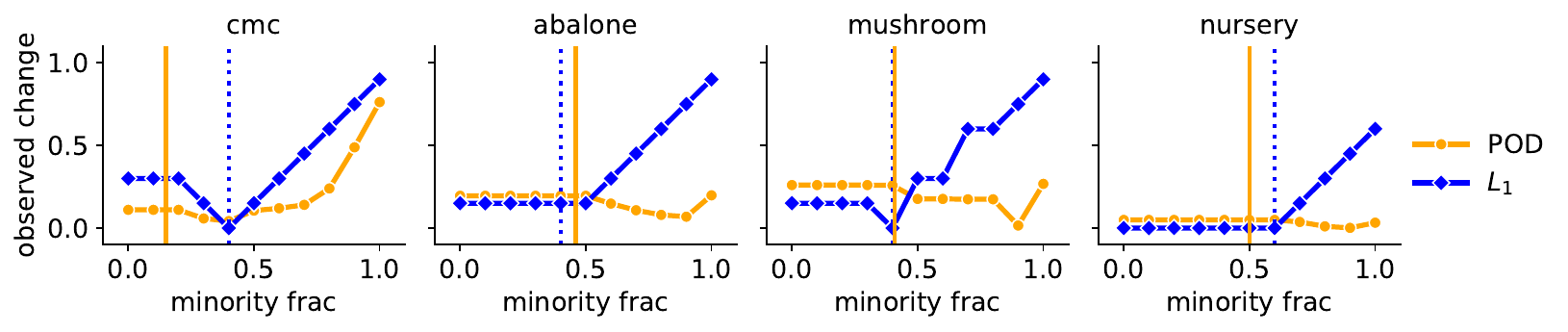}
\caption{\label{fig:experiments:pod2}
Price of diversity for moderate size datasets ($k=10$). }
\end{figure}

In Figure~\ref{fig:experiments:l1}, we report the $L_1$ measure as a function of the increase in
the minority fraction. Note that we enforce a restriction that the ratio of
minority nodes should be at least the minority fraction, however, the ratio of
facilities chosen from the minority group can be more than the minority fraction
enforced. In this experiment we measure the change in the type of facilities
chosen. We observe more variance in $L_1$ score among the independent
runs when the minority fraction of the solution is less than the minority
fraction of the dataset. This shows that the algorithm has more flexibility to
choose the different type of facilities. 
In Figure~\ref{fig:experiments:pod2} we report \POD and
$L_1$ measure for moderate size datasets.

\begin{figure}[t]
\includegraphics[width=1.0\linewidth]{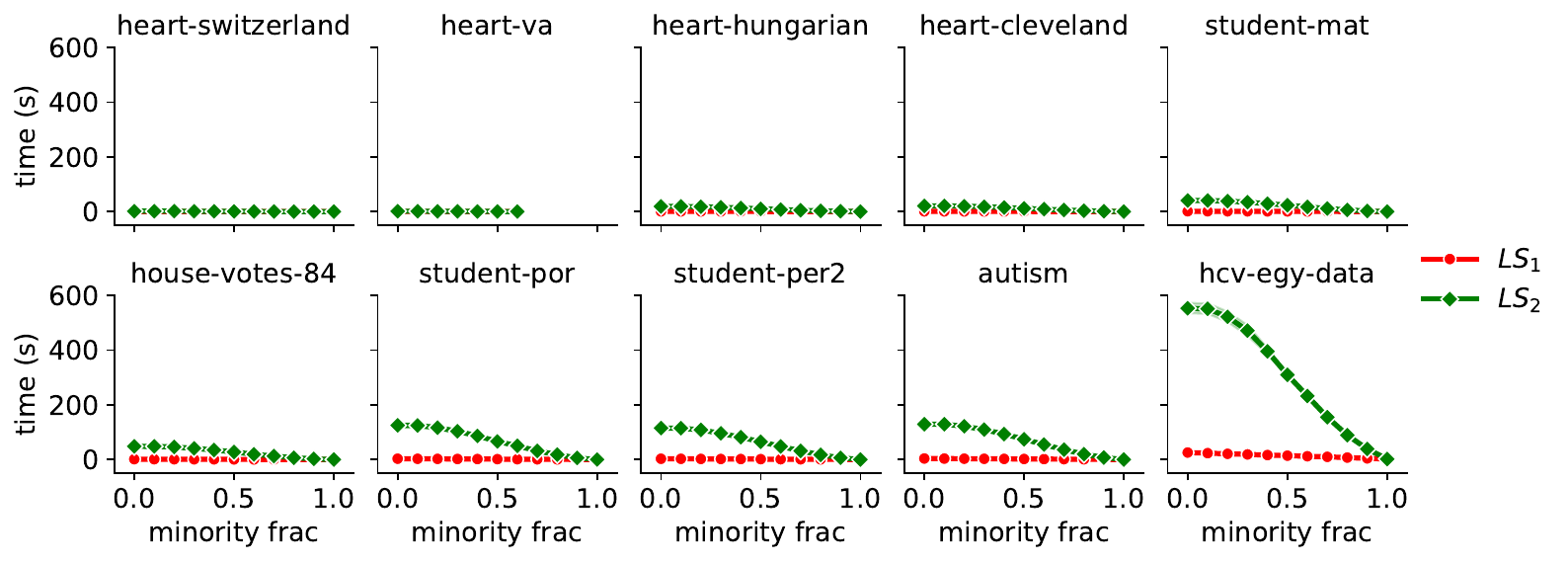}
\caption{\label{fig:experiments:runtime}
Running Time ($k=10$). }
\end{figure}

\spara{Running time.}
In Figure~\ref{fig:experiments:runtime} we report the running time of \lsone and
\lstwo as a
function of the minority fraction. For small datasets we observe no significant
change in the running time of \lsone and \lstwo. However, the dataset size
has a significant impact on running time of \lstwo. For instance in the {\tt
hcv-egy-data} dataset, for $k=10$ and minority fraction $0.1$, 
\lstwo is $300$ times slower than \lsone. This is expected, as
the algorithm considers a quadratic number of replacements per
iteration. Despite this increase in time, there is no significant
improvement in the cost of the solution obtained, as observed in
Figure~\ref{fig:experiments:pod}.
This makes \lsone our method of choice in problem instances where the facility groups are disjoint.

\begin{figure}[t]
\includegraphics[width=1.0\linewidth]{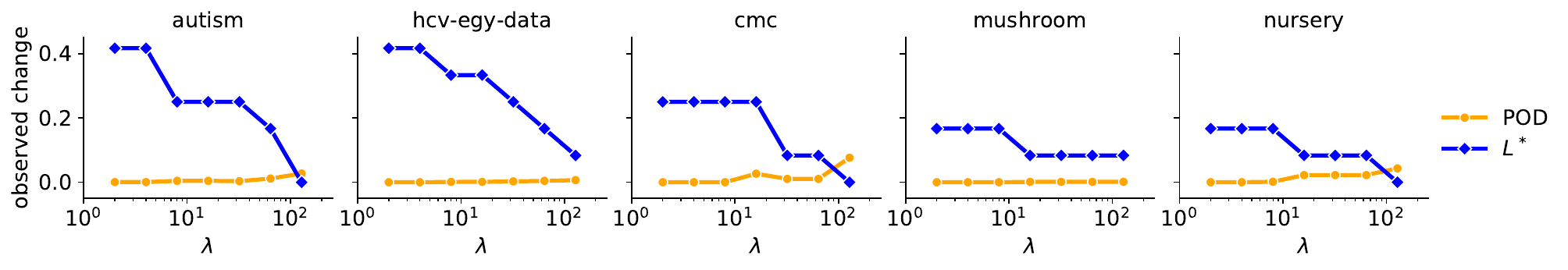}
\caption{\label{fig:experiments:tractable}
Price of diversity for intersecting facility types ($k=10$).}
\end{figure}

\spara{Relaxed objective.} In our final set of experiments we study the
behavior of the \lszero local-search heuristic with the relaxed objective
function of Equation~(\ref{eq:relaxed-cost:2}). 
In Figure~\ref{fig:experiments:tractable} we report price
of diversity (\POD) and the fraction of violations of representation constraints
$L^*$ for each value of $\lambda = \{2^1,\dots,2^7\}$. For each dataset we
choose four protected attributes to obtain intersecting facility types, and
perform experiments with $k=10$ and the representation constraints set
$R=\{3,3,3,3\}$. The value of $L^*$ measures the fraction of violations of the
representation constraints i.e, 
$L^* = \frac{\sum_{i \in [t]} \min{(0, |S\cap F_i|-r_i))}}{\sum_{i \in [t]} r_i}$.
With the increase in the value of $\lambda$ the value of $L^*$ decreases and the
value of \POD increases, as expected.
However, the increase in \POD is very small and in all cases it is 
possible to find solutions where both \POD and  $L^*$ are very close to zero, 
that is, 
solutions that have very few constraint violations and their clustering cost is almost as low 
as in the unconstrained version.

\section{Conclusion}
\label{section:conclusion}

We introduce a novel formulation of diversity-aware clustering,
which ensures fairness by avoiding under-representation of the
cluster centers, where the cluster centers belong in different
facility groups. We show that the general variant of the problem where facility groups
overlap is \np-hard, fixed-parameter intractable, and inapproximable to any
multiplicative factor. Despite such negative results we show that the variant
of the problem with disjoint facility types can be approximated efficiently.
We also present heuristic algorithms that practically solve real-world
problem instances and empirically evaluated the proposed solutions using an
extensive set of experiments. The main open problem left is to improve
the run-time complexity of the approximation algorithm, 
in the setting of disjoint groups and $t>2$, 
so that it does not use repeated calls to a linear program.
Additionally, it would be interesting to devise FPT algorithms
for obtaining exact solutions, again in the case of disjoint groups.

%
%
%
\bibliographystyle{splncs04}
\bibliography{sources-short}









\section{Proofs}
\label{sup:proofs}

\subsection{Proof of Theorem~\ref{theorem:hardness:approx2}}
\begin{proof}
Given an instance $(G, k)$ of the \vertexcover problem such that $G=(V,E)$,
$k \leq |V|$, we construct an \divkmedian problem instance
$(C, \pc{F}, \mc{F}, R, k', d)$ as follows:
($i$) $C=V$, ($ii$) $\pc{F}=V$,
($iii$) for each edge $\{u,v\} \in E$ we construct a set
$F_i = \{u, v\}$, and $\mc{F} = \{F_1,\dots,F_m\}$,
($iv$) $R = \{1^m\}$, ($v$) $k' = k$ and 
($vi$) $d(u,v) = 1$ for $(u,v) \in C \times F$.

Let $S \subseteq V$ be a solution for the \vertexcover problem. From the
construction it is clear that $|S \cap F_i| \geq 1$ for each $F_i \in
\mathcal{F}$ since each $F_i$ is a set of vertices in an edge. $|S| \leq k$ so
$S$ is also a solution for the \divkmedian problem. The argument for the other
direction is analagous to the previous arguments.

This establishes that if the \divkmedian problem has an
algorithm with polynomial time and any approximation factor then we can solve
the \vertexcover problem in polynomial time, which is most likely not possible
assuming $\p \neq \np$.\qed
\end{proof}

\subsection{Proof of Theorem~\ref{theorem:hardness:approx3}}
\begin{proof}
By the seminal work of Bartal~\cite{bartal1996probabilistic},
a metric instance of the \divkmedian problem can be embedded into a tree metric with
at most $\log(|C|+|\pazocal{F}|)$-factor change in distance.
If there exists a polynomial-time algorithm to approximate the \divkmedian problem on a tree metric
within factor $c$, then there exists a $c\,\log(|C|+|\pazocal{F}|)$-factor approximation algorithm
for any metric distance measure.
However, the existence of such an algorithm contradicts our inapproximability result in
Theorem~\ref{theorem:hardness:approx}.
So the \divkmedian problem is \np-hard to approximate to any multiplicative factor
even if the underlying metric space is a tree.
%
\qed
\end{proof}

\section{Baseline}
\label{sup:baseline}
As baseline we experimented with ($i$) exhaustive enumeration, ($ii$) linear
program solvers, and ($iii$) vanilla local-search (\lszero).
In {\em exhaustive enumeration}, we compute the cost for each $S \subseteq
\pazocal{F}$, $|S|=k$ if the constraints in $|S \cap F_i| \geq r_i$ are
satisfied and pick the solution with minimum cost. The algorithm has a
complexity of $\bigO(n^k)$ and do not practically scale to even modest size
datasets with $n=100$ datapoints and $k=4$.
Next we formulated the \divkmedian problem as a linear program to obtain a lower
bound on the cost of the solution and solved using a range of LP solvers such as
gurobi, CVXOPT and SCIP. However, all solvers failed to solve the problem
instances with modest size inputs with $n > 200$ datapoints and
$k=4$.
Finally we used a local-search algorithm with no cardinality constraints. We
call this variant of the algorithm vanilla local-search (\lszero). For each
dataset we execute $10$ iterations of \lszero with random initial assignments to
obtain a solution with the minimum cost $\ell^*$ among the independent
executions.  The cost obtained from \lszero algorithm is a $5$-approximation for
the $k$-median problem.




\end{document}